\documentclass[italian,english]{article}
\usepackage[T1]{fontenc}
\usepackage[latin1]{inputenc}
\usepackage{color}
\usepackage{graphicx}
\usepackage{amssymb}

\makeatletter

 \newcommand{\lyxaddress}[1]{
   \par {\raggedright #1 
   \vspace{1.4em}
   \noindent\par}
 }

\usepackage{babel}
\makeatother
\begin{document}

\title{\textbf{The importance of the ''magnetic'' components of gravitational
waves in the response functions of interferometers}}

\author{\textbf{Christian Corda}}

\maketitle

\lyxaddress{\begin{center}INFN - Sezione di Pisa and Università di Pisa, Largo
Pontecorvo 1, I - 56127 PISA, Italy\end{center}}

\lyxaddress{\begin{center}\textit{E-mail address:} \textcolor{blue}{christian.corda@ego-gw.it} \end{center}}

\begin{abstract}
With an enlighting analysis, Baskaran and Grishchuk have recently
shown the presence and importance of the so-called {}``magnetic''
components of gravitational waves (GWs), which have to be taken into
account in the context of the total response functions of interferometers
for GWs propagating from arbitrary directions. In this paper more
detailed angular and frequency dependences of the response functions
for the magnetic components are given in the approximation of wavelength
much larger than the linear dimensions of the interferometer, with
a specific application to the parameters of the LIGO and Virgo interferometers.
The results of this paper agree with the work of Baskaran and Grishchuk,
in which it has been shown that the identification of {}``electric''
and {}``magnetic'' contributions is unambiguous in the long-wavelength
approximation. At the end of this paper the angular and frequency
dependences of the total response functions of the LIGO and Virgo
interferometers are given. In the high-frequency regime the division
on {}``electric'' and {}``magnetic'' components becomes ambiguous,
thus the full theory of gravitational waves has to be used. Our results
are consistent with the ones of Baskaran and Grishchuk in this case
too.
\end{abstract}

\lyxaddress{PACS numbers: 04.80.Nn, 04.80.-y, 04.25.Nx}

\section{Introduction}

The design and construction of a number of sensitive detectors for
GWs is underway today. There are some laser interferometers like the
VIRGO detector, being built in Cascina, near Pisa by a joint Italian-French
collaboration \cite{key-1,key-2}, the GEO 600 detector, being built
in Hannover, Germany by a joint Anglo-Germany collaboration \cite{key-3,key-4},
the two LIGO detectors, being built in the United States (one in Hanford,
Washington and the other in Livingston, Louisiana) by a joint Caltech-Mit
collaboration \cite{key-5,key-6}, and the TAMA 300 detector, being
built near Tokyo, Japan \cite{key-7,key-8}. There are many bar detectors
currently in operation too, and several interferometers and bars are
in a phase of planning and proposal stages.

The results of these detectors will have a fundamental impact on astrophysics
and gravitation physics. There will be lots of experimental data to
be analyzed, and theorists will be forced to interact with lots of
experiments and data analysts to extract the physics from the data
stream.

Detectors for GWs will also be important to confirm or ruling out
the physical consistency of General Relativity or of any other theory
of gravitation \cite{key-9,key-10,key-11,key-12}. This is because,
in the context of Extended Theories of Gravity, some differences between
General Relativity and the others theories can be pointed out starting
by the linearized theory of gravity \cite{key-9,key-10,key-12}. 

Baskaran and Grishchuk have recently shown the presence and importance
of the so-called {}``magnetic'' components of GWs, which have to
be taken into account in the context of the total response functions
(angular patterns) of interferometers for GWs propagating from arbitrary
directions \cite{key-13}. In this paper more detailed angular and
frequency dependences of the response functions for the magnetic components
are given in the approximation of wavelength much larger than the
linear dimensions of the interferometer, with a specific application
to the parameters of the LIGO and Virgo interferometers. The results
of this paper agree with the work of \cite{key-13} in which it has
been shown that the identification of {}``electric'' and {}``magnetic''
contributions is unambiguous in the long-wavelength approximation.
At the end of this paper the angular and frequency dependences of
the total response functions of the LIGO and Virgo interferometers
are given. In the high-frequency regime the division on {}``electric''
and {}``magnetic'' components becomes ambiguous, thus the full theory
of gravitational waves has to be used \cite{key-13}. The results
presented in this paper are consistent with the ones of \cite{key-13}
in this case too.

\section{Analysis in the frame of the local observer}

In a laboratory environment on Earth, the coordinate system in which
the space-time is locally flat is typically used \cite{key-12,key-13,key-15,key-16,key-17}
and the distance between any two points is given simply by the difference
in their coordinates in the sense of Newtonian physics. In this frame,
called the frame of the local observer, GWs manifest themselves by
exerting tidal forces on the masses (the mirror and the beam-splitter
in the case of an interferometer, see figure 1). 

\begin{figure}
\includegraphics{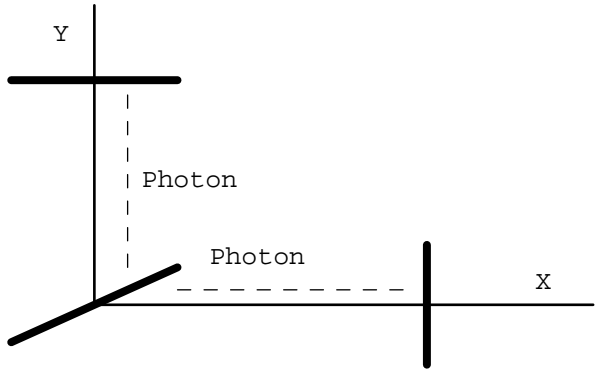}

\caption{photons can be launched from the beam-splitter to be bounced back
by the mirror}
\end{figure}
A detailed analysis of the frame of the local observer is given in
ref. \cite{key-15}, sect. 13.6. Here only the more important features
of this frame are pointed out:

the time coordinate $x_{0}$ is the proper time of the observer O;

spatial axes are centered in O;

in the special case of zero acceleration and zero rotation the spatial
coordinates $x_{j}$ are the proper distances along the axes and the
frame of the local observer reduces to a local Lorentz frame: in this
case the line element reads 

\begin{equation}
ds^{2}=-(dx^{0})^{2}+\delta_{ij}dx^{i}dx^{j}+O(|x^{j}|^{2})dx^{\alpha}dx^{\beta};\label{eq: metrica local lorentz}\end{equation}

the effect of GWs on test masses is described by the equation for
geodesic deviation in this frame

\begin{equation}
\ddot{x^{i}}=-\widetilde{R}_{0k0}^{i}x^{k},\label{eq: deviazione geodetiche}\end{equation}
where $\widetilde{R}_{0k0}^{i}$ are the components of the linearized
Riemann tensor \cite{key-15}. 

Recently the presence and importance of the so-called magnetic components
of GWs have been shown by Baskaran and Grishchuk that computed the
correspondent detector patterns in the low-frequencies approximation
\cite{key-13}. Actually,  a more detailed angular and frequency dependences
of the response functions for the magnetic components can be given
in the same approximation, with a specific application to the parameters
of the LIGO and Virgo interferometers.

Before starting with the analysis of the response functions, a brief
review of Section 3 of \cite{key-13} is necessary to understand the
importance of the {}``magnetic'' components of GWs. In this paper
we use different notations with respect to the ones used in \cite{key-13}.
We work with $G=1$, $c=1$ and $\hbar=1$ and we call $h_{+}(t_{tt}+z_{tt})$
and $h_{\times}(t_{tt}+z_{tt})$ the weak perturbations due to the
$+$ and the $\times$ polarizations which are expressed in terms
of synchronous coordinates $t_{tt},x_{tt},y_{tt},z_{tt}$ in the transverse-traceless
(TT) gauge. In this way, the most general GW propagating in the $z_{tt}$
direction can be written in terms of plane monochromatic waves \cite{key-15,key-16,key-17,key-18}

\begin{equation}
\begin{array}{c}
h_{\mu\nu}(t_{tt}+z_{tt})=h_{+}(t_{tt}+z_{tt})e_{\mu\nu}^{(+)}+h_{\times}(t_{tt}+z_{tt})e_{\mu\nu}^{(\times)}=\\
\\=h_{+0}\exp i\omega(t_{tt}+z_{tt})e_{\mu\nu}^{(+)}+h_{\times0}\exp i\omega(t_{tt}+z_{tt})e_{\mu\nu}^{(\times)},\end{array}\label{eq: onda generale}\end{equation}

and the correspondent line element will be

\begin{equation}
ds^{2}=dt_{tt}^{2}-dz_{tt}^{2}-(1+h_{+})dx_{tt}^{2}-(1-h_{+})dy_{tt}^{2}-2h_{\times}dx_{tt}dx_{tt}.\label{eq: metrica TT totale}\end{equation}

The wordlines $x_{tt},y_{tt},z_{tt}=const$ are timelike geodesics
representing the histories of free test masses \cite{key-15,key-17}.
The coordinate transformation $x^{\alpha}=x^{\alpha}(x_{tt}^{\beta})$
from the TT coordinates to the frame of the local observer is \cite{key-13,key-19}

\begin{equation}
\begin{array}{c}
t=t_{tt}+\frac{1}{4}(x_{tt}^{2}-y_{tt}^{2})\dot{h}_{+}-\frac{1}{2}x_{tt}y_{tt}\dot{h}_{\times}\\
\\x=x_{tt}+\frac{1}{2}x_{tt}h_{+}-\frac{1}{2}y_{tt}h_{\times}+\frac{1}{2}x_{tt}z_{tt}\dot{h}_{+}-\frac{1}{2}y_{tt}z_{tt}\dot{h}_{\times}\\
\\y=y_{tt}+\frac{1}{2}y_{tt}h_{+}-\frac{1}{2}x_{tt}h_{\times}+\frac{1}{2}y_{tt}z_{tt}\dot{h}_{+}-\frac{1}{2}x_{tt}z_{tt}\dot{h}_{\times}\\
\\z=z_{tt}-\frac{1}{4}(x_{tt}^{2}-y_{tt}^{2})\dot{h}_{+}+\frac{1}{2}x_{tt}y_{tt}\dot{h}_{\times},\end{array}\label{eq: trasf. coord.}\end{equation}

where it is $\dot{h}_{+}\equiv\frac{\partial h_{+}}{\partial t}$
and $\dot{h}_{\times}\equiv\frac{\partial h_{\times}}{\partial t}$.
The coefficients of this transformation (components of the metric
and its first time derivative) are taken along the central wordline
of the local observer \cite{key-13,key-14,key-19}. In refs. \cite{key-13,key-19}
it has been shown that the linear and quadratic terms, as powers of
$x_{tt}^{\alpha}$, are unambiguously determined by the conditions
of the frame of the local observer, while the cubic and higher-order
corrections are not determined by these conditions. Thus, at high-frequencies,
the expansion in terms of higher-order corrections breaks down \cite{key-13,key-14}. 

Considering a free mass riding on a timelike geodesic ($x=l_{1}$,
$y=l_{2},$ $z=l_{3}$) \cite{key-13}, eqs. (\ref{eq: trasf. coord.})
define the motion of this mass with respect to the introduced frame
of the local observer. In concrete terms one gets\begin{equation}
\begin{array}{c}
x(t)=l_{1}+\frac{1}{2}[l_{1}h_{+}(t)-l_{2}h_{\times}(t)]+\frac{1}{2}l_{1}l_{3}\dot{h}_{+}(t)+\frac{1}{2}l_{2}l_{3}\dot{h}_{\times}(t)\\
\\y(t)=l_{2}-\frac{1}{2}[l_{2}h_{+}(t)+l_{1}h_{\times}(t)]-\frac{1}{2}l_{2}l_{3}\dot{h}_{+}(t)+\frac{1}{2}l_{1}l_{3}\dot{h}_{\times}(t)\\
\\z(t)=l_{3}-\frac{1}{4[}(l_{1}^{2}-l_{2}^{2})\dot{h}_{+}(t)+2l_{1}l_{2}\dot{h}_{\times}(t),\end{array}\label{eq: Grishuk 0}\end{equation}
which are exactly eqs. (13) of \cite{key-13} rewritten using our
notation. In absence of GWs the position of the mass is $(l_{1},l_{2},l_{3}).$
The effect of the GW is to drive the mass to have oscillations. Thus,
in general, from eqs. (\ref{eq: Grishuk 0}) all three components
of motion are present \cite{key-13}.

Neglecting the terms with $\dot{h}_{+}$ and $\dot{h}_{\times}$ in
eqs. (\ref{eq: Grishuk 0}), the {}``traditional'' equations for
the mass motion are obtained \cite{key-15,key-17,key-18}:\begin{equation}
\begin{array}{c}
x(t)=l_{1}+\frac{1}{2}[l_{1}h_{+}(t)-l_{2}h_{\times}(t)]\\
\\y(t)=l_{2}-\frac{1}{2}[l_{2}h_{+}(t)+l_{1}h_{\times}(t)]\\
\\z(t)=l_{3}.\end{array}\label{eq: traditional}\end{equation}

Clearly, this is the analogue of the electric component of motion
in electrodynamics \cite{key-13}, while equations\begin{equation}
\begin{array}{c}
x(t)=l_{1}+\frac{1}{2}l_{1}l_{3}\dot{h}_{+}(t)+\frac{1}{2}l_{2}l_{3}\dot{h}_{\times}(t)\\
\\y(t)=l_{2}-\frac{1}{2}l_{2}l_{3}\dot{h}_{+}(t)+\frac{1}{2}l_{1}l_{3}\dot{h}_{\times}(t)\\
\\z(t)=l_{3}-\frac{1}{4[}(l_{1}^{2}-l_{2}^{2})\dot{h}_{+}(t)+2l_{1}l_{2}\dot{h}_{\times}(t),\end{array}\label{eq: news}\end{equation}

are the analogue of the magnetic component of motion. One could think
that the presence of these magnetic components is a {}``frame artefact''
due to the transformation (\ref{eq: trasf. coord.}), but in Section
4 of \cite{key-13} eqs. (\ref{eq: Grishuk 0}) have been directly
obtained from the geodesic deviation equation too, thus the magnetic
components have a real physical significance. The fundamental point
of \cite{key-13} is that the magnetic components become important
when the frequency of the wave increases (Section 3 of \cite{key-13}),
but only in the low-frequency regime. This can be understood directly
from eqs. (\ref{eq: Grishuk 0}). In fact, using eqs. (\ref{eq: onda generale})
and (\ref{eq: trasf. coord.}), eqs. (\ref{eq: Grishuk 0}) become\begin{equation}
\begin{array}{c}
x(t)=l_{1}+\frac{1}{2}[l_{1}h_{+}(t)-l_{2}h_{\times}(t)]+\frac{1}{2}l_{1}l_{3}\omega h_{+}(t)+\frac{1}{2}l_{2}l_{3}\omega h_{\times}(t)\\
\\y(t)=l_{2}-\frac{1}{2}[l_{2}h_{+}(t)+l_{1}h_{\times}(t)]-\frac{1}{2}l_{2}l_{3}\omega h_{+}(t)+\frac{1}{2}l_{1}l_{3}\omega h_{\times}(t)\\
\\z(t)=l_{3}-\frac{1}{4[}(l_{1}^{2}-l_{2}^{2})\omega h_{+}(t)+2l_{1}l_{2}\omega h_{\times}(t).\end{array}\label{eq: Grishuk 01}\end{equation}

Thus the terms with $\dot{h}_{+}$ and $\dot{h}_{\times}$ in eqs.
(\ref{eq: Grishuk 0}) can be neglected only when the wavelength goes
to infinity \cite{key-13}, while, at high-frequencies, the expansion
in terms of $\omega l_{i}l_{j}$ corrections, with $i,j=1,2,3,$ breaks
down \cite{key-13,key-14}. 

Now, let us compute the total response functions of interferometers
for the magnetic components.

Equations (\ref{eq: Grishuk 0}), that represent the coordinates of
the mirror of the interferometer in presence of a GW in the frame
of the local observer, can be rewritten for the pure magnetic component
of the $+$ polarization as

\begin{equation}
\begin{array}{c}
x(t)=l_{1}+\frac{1}{2}l_{1}l_{3}\dot{h}_{+}(t)\\
\\y(t)=l_{2}-\frac{1}{2}l_{2}l_{3}\dot{h}_{+}(t)\\
\\z(t)=l_{3}-\frac{1}{4}(l_{1}^{2}-l_{2}^{2})\dot{h}_{+}(t),\end{array}\label{eq: Grishuk 1}\end{equation}

where $l_{1},l_{2}\textrm{ }and\textrm{ }\textrm{ }l_{3}$ are the
unperturbed coordinates of the mirror. 

To compute the response functions for an arbitrary propagating direction
of the GW, we recall that the arms of the interferometer are in general
in the $\overrightarrow{u}$ and $\overrightarrow{v}$ directions,
while the $x,y,z$ frame is adapted to the propagating GW (i.e. the
observer is assumed located in the position of the beam splitter).
Then a spatial rotation of the coordinate system has to be performed:

\begin{equation}
\begin{array}{ccc}
u & = & -x\cos\theta\cos\phi+y\sin\phi+z\sin\theta\cos\phi\\
\\v & = & -x\cos\theta\sin\phi-y\cos\phi+z\sin\theta\sin\phi\\
\\w & = & x\sin\theta+z\cos\theta,\end{array}\label{eq: rotazione}\end{equation}

or, in terms of the $x,y,z$ frame:

\begin{equation}
\begin{array}{ccc}
x & = & -u\cos\theta\cos\phi-v\cos\theta\sin\phi+w\sin\theta\\
\\y & = & u\sin\phi-v\cos\phi\\
\\z & = & u\sin\theta\cos\phi+v\sin\theta\sin\phi+w\cos\theta.\end{array}\label{eq: rotazione 2}\end{equation}

In this way the GW is propagating from an arbitrary direction $\overrightarrow{r}$
to the interferometer (see figure 2). Because the mirror of eqs. (\ref{eq: Grishuk 1})
is situated in the $u$ direction, using eqs. (\ref{eq: Grishuk 1}),
(\ref{eq: rotazione}) and (\ref{eq: rotazione 2}) the $u$ coordinate
of the mirror is given by

\begin{figure}
\includegraphics{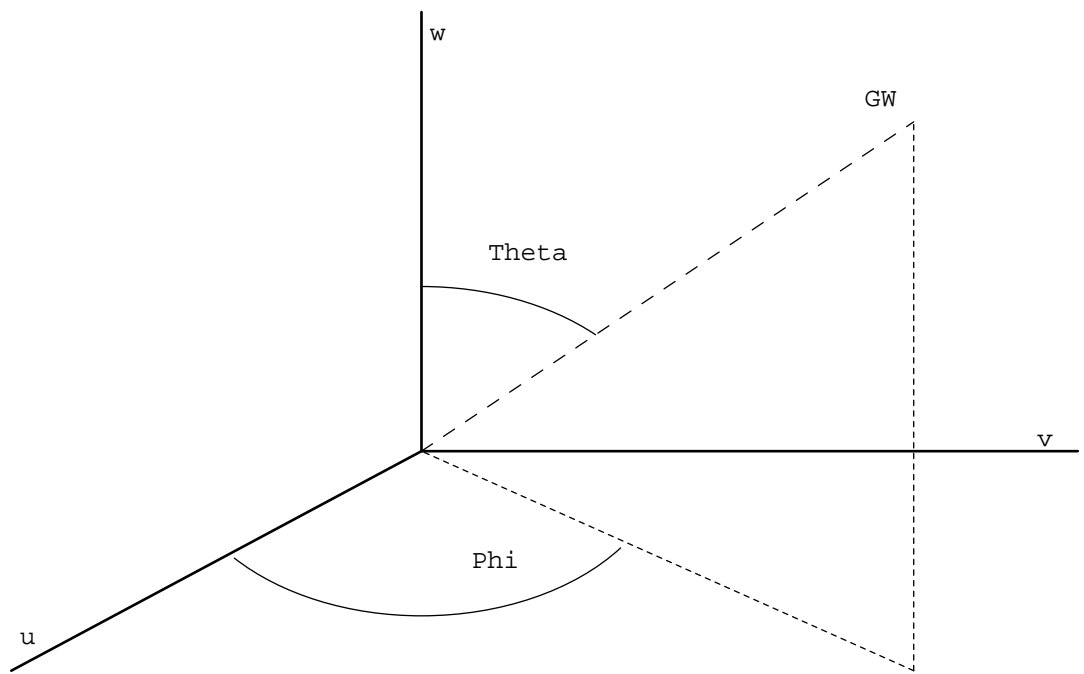}

\caption{a GW propagating from an arbitrary direction}
\end{figure}

\begin{equation}
u=L+\frac{1}{4}L^{2}A\dot{h}_{+}(t),\label{eq: du}\end{equation}

where \begin{equation}
A\equiv\sin\theta\cos\phi(\cos^{2}\theta\cos^{2}\phi-\sin^{2}\phi)\label{eq: A}\end{equation}

and $L=\sqrt{l_{1}^{2}+l_{2}^{2}+l_{3}^{2}}$ is the length of the
interferometer arms.

The computation for the $v$ arm is similar to the one above. Using
eqs. (\ref{eq: Grishuk 1}), (\ref{eq: rotazione}) and (\ref{eq: rotazione 2}),
the coordinate of the mirror in the $v$ arm is:

\begin{equation}
v=L+\frac{1}{4}L^{2}B\dot{h}_{+}(t),\label{eq: dv}\end{equation}

where\begin{equation}
B\equiv\sin\theta\sin\phi(\cos^{2}\theta\cos^{2}\phi-\sin^{2}\phi).\label{eq: B}\end{equation}

\section{The response function of an interferometer for the magnetic contribution
of the $+$ polarization}

Equations (\ref{eq: du}) and (\ref{eq: dv}) represent the distance
of the two mirrors of the interferometer from the beam-splitter in
presence of the GW (i.e. only the contribution of the magnetic component
of the $+$ polarization of the GW is taken into account). They represent
particular cases of the more general form given in eq. (33) of \cite{key-13}.

A {}``signal'' can also be defined in the time domain (i.e. $T=L$
in our notation):

\begin{equation}
\frac{\delta T(t)}{T}\equiv\frac{u-v}{L}=\frac{1}{4}L(A-B)\dot{h}_{+}(t).\label{eq: signal piu}\end{equation}

The quantity (\ref{eq: signal piu}) can be computed in the frequency
domain using the Fourier transform of $h_{+}$, defined by

\begin{equation}
\tilde{h}_{+}(\omega)=\int_{-\infty}^{\infty}dth_{+}(t)\exp(i\omega t),\label{eq: trasformata di fourier}\end{equation}
obtaining

\[
\frac{\tilde{\delta}T(\omega)}{T}=H_{magn}^{+}(\omega)\tilde{h}_{+}(\omega),\]

where the function

\begin{equation}
\begin{array}{c}
H_{magn}^{+}(\omega)=-\frac{1}{8}i\omega L(A-B)=\\
\\=-\frac{1}{4}i\omega L\sin\theta[(\cos^{2}\theta+\sin2\phi\frac{1+\cos^{2}\theta}{2})](\cos\phi-\sin\phi)\end{array}\label{eq: risposta totale}\end{equation}

is the total response function of the interferometer for the magnetic
component of the $+$ polarization, in perfect agreement with the
result of Baskaran and Grishchuk (eqs. 46 and 49 of \cite{key-13}).
In the above computation the theorem on the derivative of the Fourier
transform has been used.

In the present work the $x,y,z$ frame is the frame of the local observer
adapted to the propagating GW, while in \cite{key-13} the two frames
are not in phase (i.e. in this paper the third angle is put equal
to zero, this is not a restriction as it is known in literature, see
for example \cite{key-12}).

The absolute value of the response functions (\ref{eq: risposta totale})
of the Virgo ($L=3$Km) and LIGO ($L=4$Km) interferometers to the
magnetic component of the $+$ polarization for $\theta=\frac{\pi}{4}$
and $\phi=\frac{\pi}{3}$ are respectively shown in figures 3 and
4 in the low-frequency range $10Hz\leq f\leq100Hz$. This quantity
increases with increasing frequency.%
\begin{figure}
\includegraphics{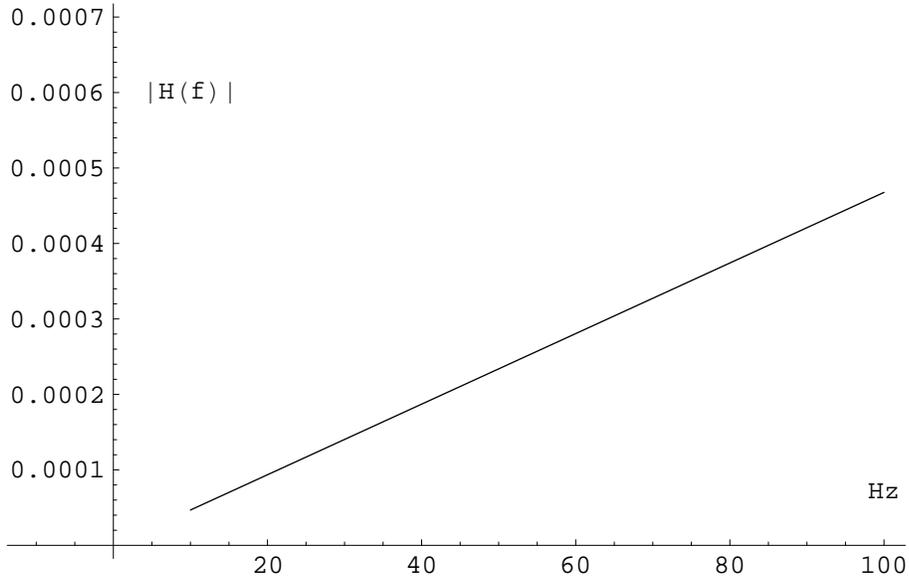}

\caption{the absolute value of the total response function of the Virgo interferometer
to the magnetic component of the $+$ polarization for $\theta=\frac{\pi}{4}$
and $\phi=\frac{\pi}{3}$ in the low-frequency range $10Hz\leq f\leq100Hz$}
\end{figure}
\begin{figure}
\includegraphics{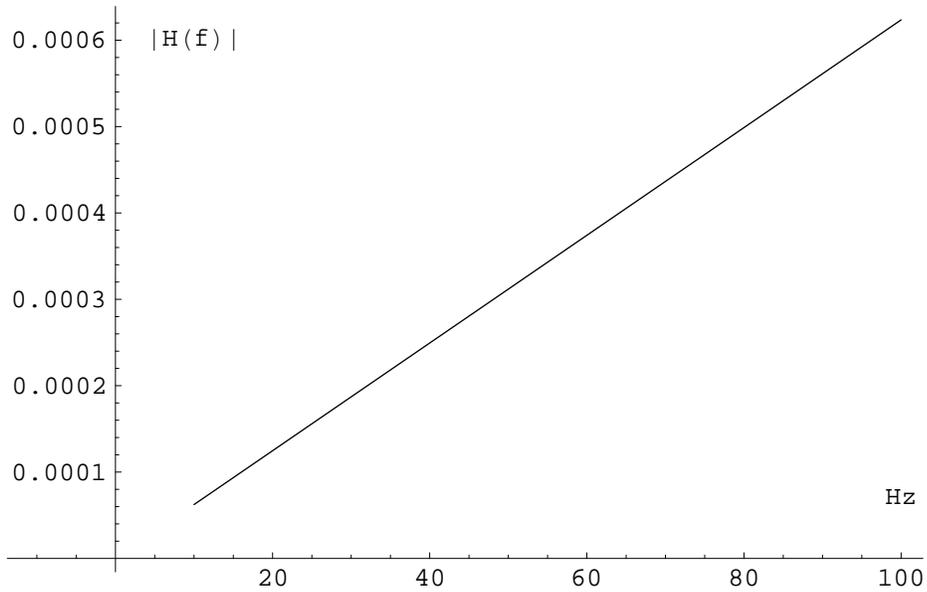}

\caption{the absolute value of the total response function of the LIGO interferometer
to the magnetic component of the $+$ polarization for $\theta=\frac{\pi}{4}$
and $\phi=\frac{\pi}{3}$ in the low- frequency range $10Hz\leq f\leq100Hz$}
\end{figure}
The angular dependences of the response function (\ref{eq: risposta totale})
of the Virgo and LIGO interferometers to the magnetic component of
the $+$ polarization for $f=100Hz$ are shown in figures 5 and 6. 

\begin{figure}
\includegraphics{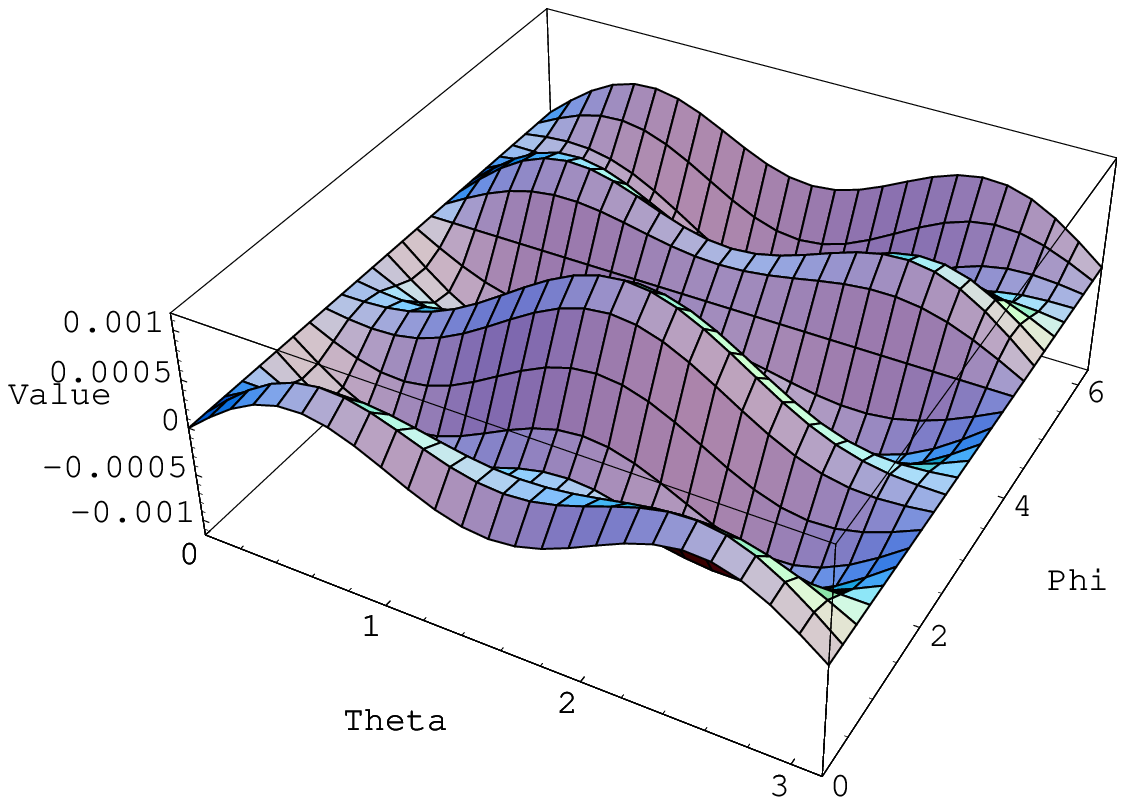}

\caption{the angular dependence of the response function of the Virgo interferometer
to the magnetic component of the $+$ polarization for $f=100Hz$}
\end{figure}
\begin{figure}
\includegraphics{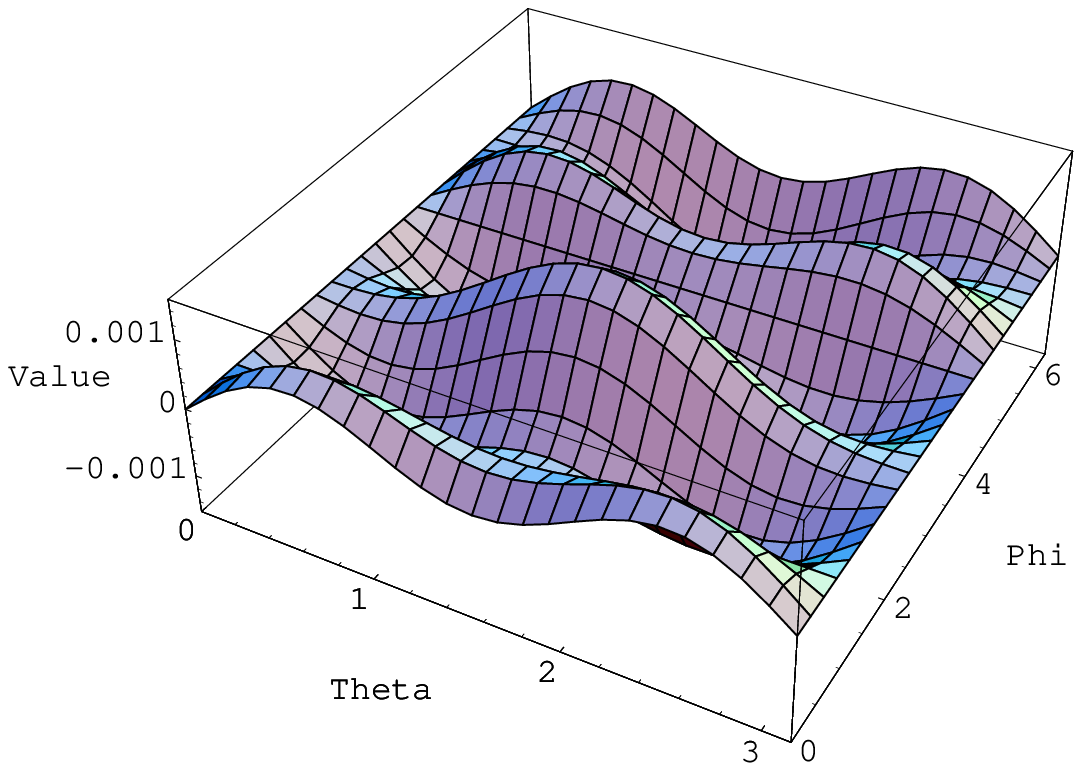}

\caption{the angular dependence of the response function of the LIGO interferometer
to the magnetic component of the $+$ polarization for $f=100Hz$}
\end{figure}

\section{Analysis for the $\times$ polarization}

The analysis can be generalized for the magnetic component of the
$\times$ polarization too. In this case, equations (\ref{eq: Grishuk 0})
can be rewritten for the pure magnetic component of the $\times$
polarization as

\begin{equation}
\begin{array}{c}
x(t+z)=l_{1}+\frac{1}{2}l_{2}l_{3}\dot{h}_{\times}(t+z)\\
\\y(t+z)=l_{2}+\frac{1}{2}l_{1}l_{3}\dot{h}_{\times}(t+z)\\
\\z(t+z)=l_{3}-\frac{1}{2}l_{1}l_{2}\dot{h}_{\times}(t+z).\end{array}\label{eq: Grishuk 2}\end{equation}

Using eqs. (\ref{eq: Grishuk 2}), (\ref{eq: rotazione}) and (\ref{eq: rotazione 2}),
the $u$ coordinate of the mirror in the $u$ arm of the interferometer
is given by \begin{equation}
u=L+\frac{1}{4}L^{2}C\dot{h}_{\times}(t),\label{eq: du C}\end{equation}

where \begin{equation}
C\equiv-2\cos\theta\cos^{2}\phi\sin\theta\sin\phi,\label{eq: C}\end{equation}
while the $v$ coordinate of the mirror in the $v$ arm of the interferometer
is given by \begin{equation}
v=L+\frac{1}{4}L^{2}D\dot{h}_{\times}(t),\label{eq: dv  D}\end{equation}

where \begin{equation}
D\equiv2\cos\theta\cos\phi\sin\theta\sin^{2}\phi.\label{eq: D}\end{equation}

Thus, with an analysis similar to the one of previous Sections, it
is possible to show that the response function of the interferometer
for the magnetic component of the $\times$ polarization is\begin{equation}
\begin{array}{c}
H_{magn}^{\times}(\omega)=-i\omega T(C-D)=\\
\\=-i\omega L\sin2\phi(\cos\phi+\sin\phi)\cos\theta,\end{array}\label{eq: risposta totale 2 per}\end{equation}

in perfect agreement with the result of Baskaran and Grishchuk (eqs.
46 and 50 of \cite{key-13}). The absolute value of the total response
functions (\ref{eq: risposta totale 2 per}) of the Virgo and LIGO
interferometers to the magnetic component of the $\times$ polarization
for $\theta=\frac{\pi}{4}$ and $\phi=\frac{\pi}{3}$ are respectively
shown in figure 7 and 8 in the low- frequency range $10Hz\leq f\leq100Hz$.
This quantity increases with increasing frequency in analogy with
the case shown in previous Section for the magnetic component of the
$+$ polarization. %
\begin{figure}
\includegraphics{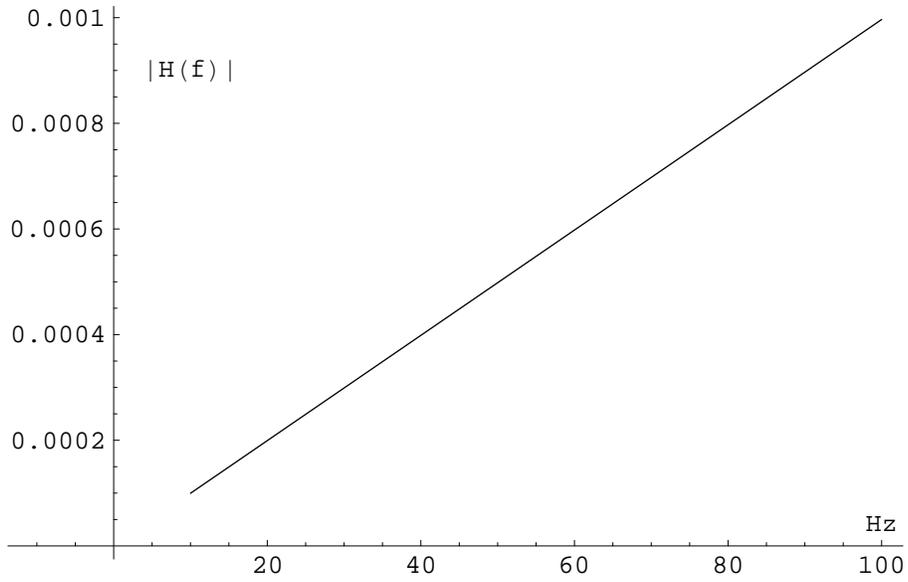}

\caption{the absolute value of the total response function of the Virgo interferometer
to the magnetic component of the $\times$ polarization for $\theta=\frac{\pi}{4}$
and $\phi=\frac{\pi}{3}$ in the low- frequency range $10Hz\leq f\leq100Hz$}
\end{figure}
\begin{figure}
\includegraphics{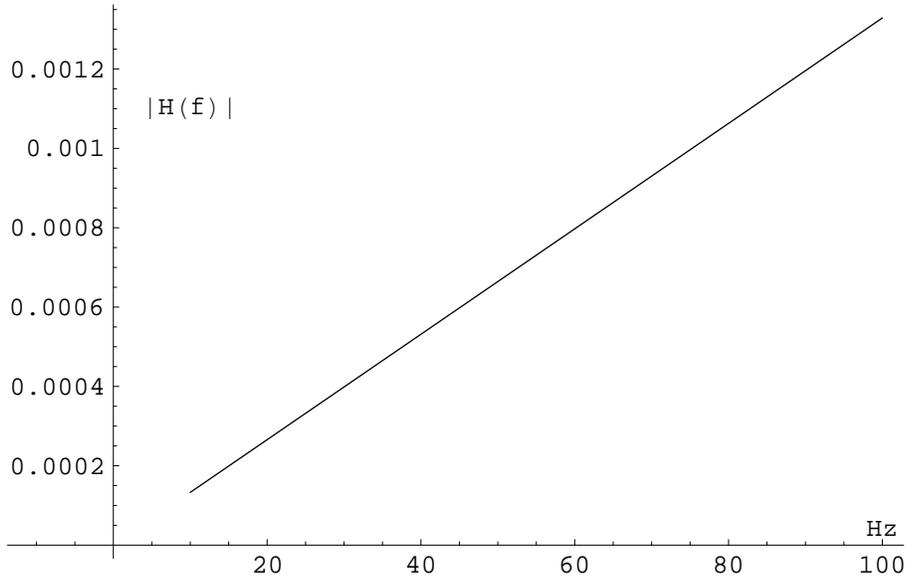}

\caption{the absolute value of the total response function of the LIGO interferometer
to the magnetic component of the $\times$ polarization for $\theta=\frac{\pi}{4}$
and $\phi=\frac{\pi}{3}$ in the low- frequency range $10Hz\leq f\leq100Hz$}
\end{figure}
 The angular dependences of the total response function (\ref{eq: risposta totale 2 per})
of the Virgo and LIGO interferometers to the magnetic component of
the $\times$ polarization for $f=100Hz$ are shown in figure 9 and
10. %
\begin{figure}
\includegraphics{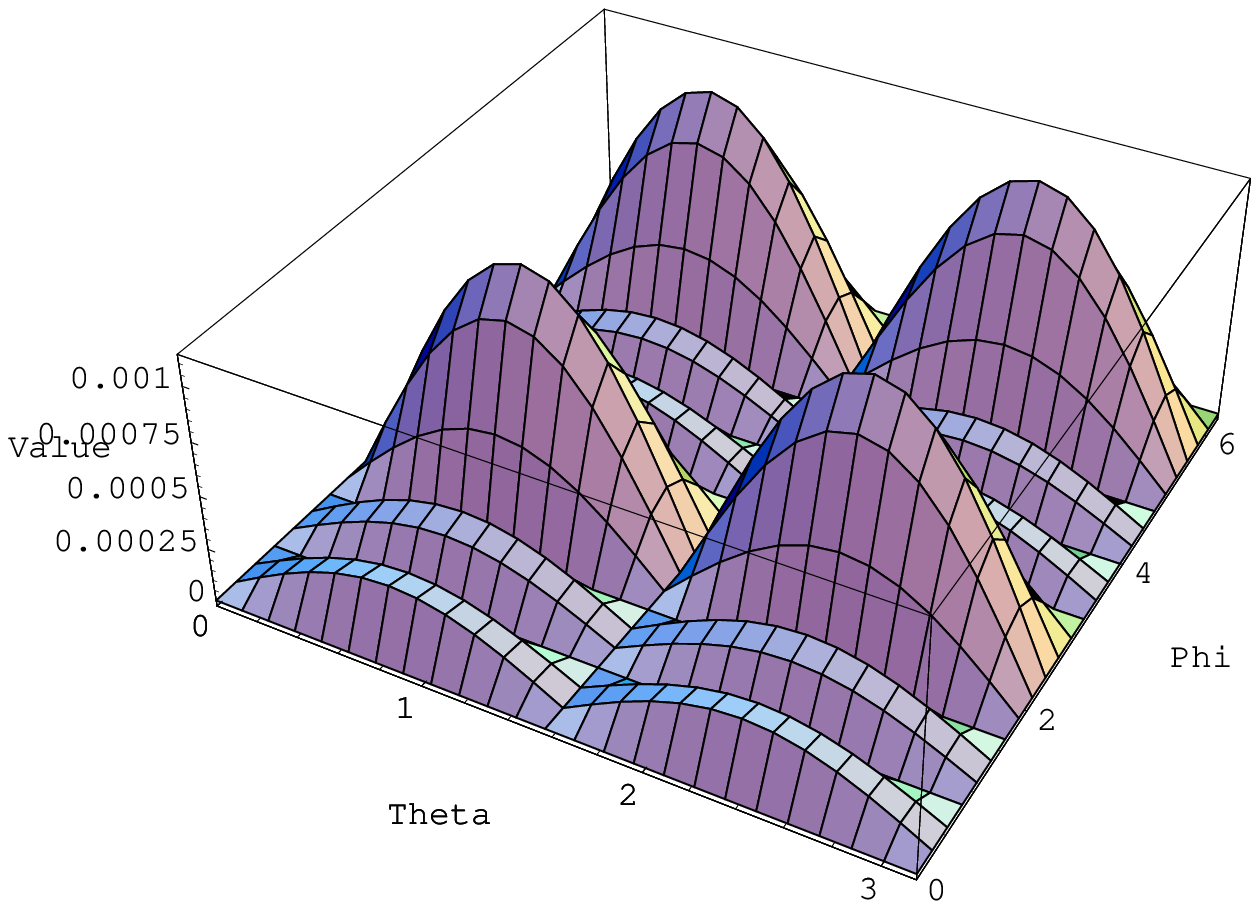}

\caption{the angular dependence of the total response function of the Virgo
interferometer to the magnetic component of the $\times$ polarization
for $f=100Hz$}
\end{figure}
\begin{figure}
\includegraphics{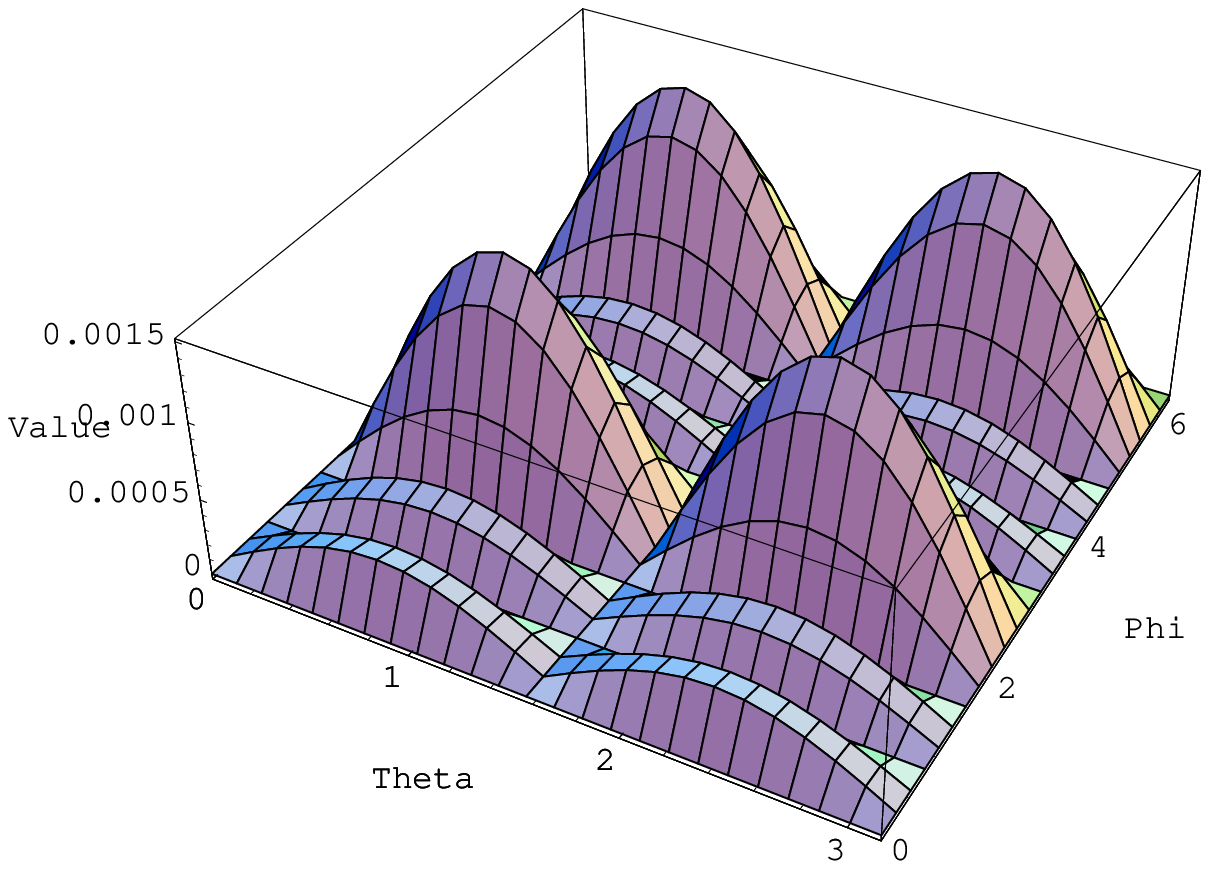}

\caption{the angular dependence of the total response function of the LIGO
interferometer to the magnetic component of the $\times$ polarization
for $f=100Hz$}
\end{figure}

\section{The total response function of interferometers in the full theory
of gravitational waves}

The low-frequencies approximation, used in previous Sections to show
that the {}``magnetic'' and {}``electric'' contributions to the
response functions can be identified without ambiguity in the longh-wavelengths
regime (see also \cite{key-13}), is sufficient only for ground based
interferometers, for which the condition $f\ll1/L$ is in general
satisfied. For space-based interferometers, for which the above condition
is not satisfied in the high-frequency portion of the sensitivity
band \cite{key-13,key-14,key-22,key-23}, the full theory of gravitational
waves has to be used.

If one removes the low-frequencies approximation, an analysis similiar
to the one used for the first time in \cite{key-16} can be used:
the so called {}``bouncing photon method''. This method has been
generalized to scalar waves, angular dependences and massive modes
of GWs in \cite{key-12}. This is also a part of the more general
problem of finding the null geodesics of light in the presence of
a weak gravitational wave \cite{key-13,key-15,key-20,key-21,key-22,key-23}.

In this section the variation of the proper distance that a photon
covers to make  a round-trip from the beam-splitter to the mirror
of an interferometer \cite{key-12,key-16} is computed with the gauge
choice (\ref{eq: metrica TT totale}). In this case one does not need
the coordinate transformation (\ref{eq: trasf. coord.}) from the
TT coordinates to the frame of the local observer (see also Section
5 of \cite{key-13}). Thus, with a treatment parallel to the one of
\cite{key-12,key-16}, the analysis is translated in the frequency
domain and the general response functions are obtained.

A special property of the TT gauge is that an inertial test mass initially
at rest in these coordinates, remains at rest throughout the entire
passage of the GW \cite{key-15,key-16,key-18}. Here we have to clarify
the use of words {}`` at rest'': we want to mean that the coordinates
of the test mass do not change in the presence of the GW. The proper
distance between the beam-splitter and the mirror of the interferometer
changes even though their coordinates remain the same \cite{key-15,key-16}.

We start from the $+$ polarization. Labelling the coordinates of
the TT gauge with $t,x,y,z$ the line element (\ref{eq: metrica TT totale})
becomes:

\begin{equation}
ds^{2}=-dt^{2}+dz^{2}+[1+h_{+}(t+z)]dx^{2}+[1+h_{+}(t+z)]dy^{2}.\label{eq: metrica polarizzazione +}\end{equation}

But the arms of the interferometer are in the $\overrightarrow{u}$
and $\overrightarrow{v}$ directions, while the $x,y,z$ frame is
the proper frame of the propagating GW. 

The coordinate transformation for the metric tensor is \cite{key-17}:

\begin{equation}
g^{ik}=\frac{\partial x^{i}}{\partial x'^{l}}\frac{\partial x^{k}}{\partial x'^{m}}g'^{lm}.\label{eq: trasformazione metrica}\end{equation}

By using eq. (\ref{eq: rotazione}), (\ref{eq: rotazione 2}) and
(\ref{eq: trasformazione metrica}), in the new rotated frame the
line element (\ref{eq: metrica polarizzazione +}) in the $\overrightarrow{u}$
direction becomes:

\begin{equation}
ds^{2}=-dt^{2}+[1+(\cos^{2}\theta\cos^{2}\phi-\sin^{2}\phi)h_{+}(t+u\sin\theta\cos\phi)]du^{2}.\label{eq: metrica + lungo u}\end{equation}

Unlike the line element in eq. 2 of ref. \cite{key-16}, where there
is a pure time dependence because of the simplest geometry, in the
line element (\ref{eq: metrica + lungo u}) both a spatial dependence
in the $u$ direction and an angular dependence appear. Thus, the
present analysis is more general than the analysis of \cite{key-16},
and similar to the one of Section 7 of \cite{key-12} for the angular
response function of the scalar component. 

A good way to analyze variations in the proper distance (time) is
by means of {}``bouncing photons'' (see \cite{key-12,key-13,key-16,key-20,key-21,key-22}
and figure 1). A photon can be launched from the beam-splitter to
be bounced back by the mirror. 

The condition for null geodesics ($ds^{2}=0$) in eq. (\ref{eq: metrica + lungo u})
gives the coordinate velocity of the photon:

\begin{equation}
v^{2}\equiv(\frac{du}{dt})^{2}=\frac{1}{[1+(\cos^{2}\theta\cos^{2}\phi-\sin^{2}\phi)h_{+}(t+u\sin\theta\cos\phi)]},\label{eq: velocita' fotone u}\end{equation}

which is a convenient quantity for calculations of the photon propagation
time between the beam-splitter and the mirror \cite{key-12,key-16}.
We recall that the beam splitter is located in the origin of the new
coordinate system (i.e. $u_{b}=0$, $v_{b}=0$, $w_{b}=0$). The coordinates
of the beam-splitter $u_{b}=0$ and of the mirror $u_{m}=L$ do not
change under the influence of the GW, thus the duration of the forward
trip can be written as

\begin{equation}
T_{1}(t)=\int_{0}^{L}\frac{du}{v(t'+u\sin\theta\cos\phi)},\label{eq: durata volo}\end{equation}

with 

\begin{center}$t'=t-(L-u)$.\end{center}

In the last equation $t'$ is the retardation time (i.e. $t$ is the
time at which the photon arrives in the position $L$, so $L-u=t-t'$).

At first order in $h_{+}$ this integral can be approximated with

\begin{equation}
T_{1}(t)=T+\frac{\cos^{2}\theta\cos^{2}\phi-\sin^{2}\phi}{2}\int_{0}^{L}h_{+}(t'+u\sin\theta\cos\phi)du,\label{eq: durata volo andata approssimata u}\end{equation}

where

\begin{center}$T=L$ \end{center}

is the transit time of the photon in absence of the GW. Similiary,
the duration of the return trip will be\begin{equation}
T_{2}(t)=T+\frac{\cos^{2}\theta\cos^{2}\phi-\sin^{2}\phi}{2}\int_{L}^{0}h_{+}(t'+u\sin\theta\cos\phi)(-du),\label{eq: durata volo ritorno approssimata u}\end{equation}

though now the retardation time is 

\begin{center}$t'=t-(u-l)$.\end{center}

The round-trip time will be the sum of $T_{2}(t)$ and $T_{1}[t-T_{2}(t)]$.
The latter can be approximated by $T_{1}(t-T)$ because the difference
between the exact and the approximate values is second order in $h_{+}$.
Then, to first order in $h_{+}$, the duration of the round-trip will
be

\begin{equation}
T_{r.t.}(t)=T_{1}(t-T)+T_{2}(t).\label{eq: durata round trip}\end{equation}

By using eqs. (\ref{eq: durata volo andata approssimata u}) and (\ref{eq: durata volo ritorno approssimata u})
one sees immediately that deviations of this round-trip time (i.e.
proper distance) from its unperturbed value are given by

\begin{equation}
\begin{array}{c}
\delta T(t)=\frac{\cos^{2}\theta\cos^{2}\phi-\sin^{2}\phi}{2}\int_{0}^{L}[h_{+}(t-2T-u(1-\sin\theta\cos\phi))+\\
\\+h_{+}(t+u(1+\sin\theta\cos\phi))]du.\end{array}\label{eq: variazione temporale in u}\end{equation}

Now, using the Fourier transform of the $+$ polarization of the field,
defined by eq. (\ref{eq: trasformata di fourier}), one obtains in
the frequency domain:

\begin{equation}
\delta\tilde{T}(\omega)=\frac{1}{2}(\cos^{2}\theta\cos^{2}\phi-\sin^{2}\phi)\tilde{H}_{u}(\omega,\theta,\phi)\tilde{h}_{+}(\omega),\label{eq: segnale in frequenza lungo u}\end{equation}

where

\begin{equation}
\begin{array}{c}
\tilde{H}_{u}(\omega,\theta,\phi)=\frac{-1+\exp(2i\omega L)}{2i\omega(1+\sin^{2}\theta\cos^{2}\phi)}+\\
\\+\frac{-\sin\theta\cos\phi((1+\exp(2i\omega L)-2\exp i\omega L(1-\sin\theta\cos\phi)))}{2i\omega(1+\sin\theta\cos^{2}\phi)}\end{array}\label{eq: fefinizione Hu}\end{equation}

and we immediately see that $\tilde{H}_{u}(\omega,\theta,\phi)\rightarrow L$
when $\omega\rightarrow0$.

Thus, the total response function of the $u$ arm of the interferometer
to the $+$ component is:

\begin{equation}
\Upsilon_{u}^{+}(\omega)=\frac{(\cos^{2}\theta\cos^{2}\phi-\sin^{2}\phi)}{2L}\tilde{H}_{u}(\omega,\theta,\phi),\label{eq: risposta + lungo u}\end{equation}

where $2L=2T$ is the round-trip time in absence of gravitational
waves.

In the same way, the line element (\ref{eq: metrica polarizzazione +})
in the $\overrightarrow{v}$ direction becomes:

\begin{equation}
ds^{2}=-dt^{2}+[1+(\cos^{2}\theta\sin^{2}\phi-\cos^{2}\phi)h_{+}(t+v\sin\theta\sin\phi)]dv^{2},\label{eq: metrica + lungo v}\end{equation}

and the response function of the $v$ arm of the interferometer to
the $+$ polarization is: 

\begin{equation}
\Upsilon_{v}^{+}(\omega)=\frac{(\cos^{2}\theta\sin^{2}\phi-\cos^{2}\phi)}{2L}\tilde{H}_{v}(\omega,\theta,\phi)\label{eq: risposta + lungo v}\end{equation}

where, now 

\begin{equation}
\begin{array}{c}
\tilde{H}_{v}(\omega,\theta,\phi)=\frac{-1+\exp(2i\omega L)}{2i\omega(1+\sin^{2}\theta\sin^{2}\phi)}+\\
\\+\frac{-\sin\theta\sin\phi((1+\exp(2i\omega L)-2\exp i\omega L(1-\sin\theta\sin\phi)))}{2i\omega(1+\sin^{2}\theta\sin^{2}\phi)},\end{array}\label{eq: fefinizione Hv}\end{equation}

with $\tilde{H}_{v}(\omega,\theta,\phi)\rightarrow L$ when $\omega\rightarrow0$.
In this case the variation of the distance (time) is\begin{equation}
\delta\tilde{T}(\omega)=\frac{1}{2}(\cos^{2}\theta\cos^{2}\phi-\cos^{2}\phi)\tilde{H}_{v}(\omega,\theta,\phi)\tilde{h}_{+}(\omega).\label{eq: segnale in frequenza lungo v}\end{equation}

From equations (\ref{eq: segnale in frequenza lungo u}) and (\ref{eq: segnale in frequenza lungo v}),
the total lengths of the two arms in presence of the $+$ polarization
of the GW and in the frequency domain are: \begin{equation}
\tilde{T}_{u}(\omega)=\frac{1}{2}(\cos^{2}\theta\cos^{2}\phi-\sin^{2}\phi)\tilde{H}_{u}(\omega,\theta,\phi)\tilde{h}_{+}(\omega)+T.\label{eq: lunghezza u}\end{equation}
\begin{equation}
\tilde{T}_{v}(\omega)=\frac{1}{2}(\cos^{2}\theta\cos^{2}\phi-\cos^{2}\phi)\tilde{H}_{v}(\omega,\theta,\phi)\tilde{h}_{+}(\omega)+T,\label{eq: lunghezza v}\end{equation}

that are particular cases of the more general equation (39) in \cite{key-13}.

Thus the total frequency-dependent response function (i.e. the detector
pattern) of an interferometer to the $+$ polarization of the GW is:

\begin{equation}
\begin{array}{c}
\tilde{H}^{+}(\omega)=\Upsilon_{u}^{+}(\omega)-\Upsilon_{v}^{+}(\omega)=\\
\\=\frac{(\cos^{2}\theta\cos^{2}\phi-\sin^{2}\phi)}{2L}\tilde{H}_{u}(\omega,\theta,\phi)+\\
\\-\frac{(\cos^{2}\theta\sin^{2}\phi-\cos^{2}\phi)}{2L}\tilde{H}_{v}(\omega,\theta,\phi)\end{array}\label{eq: risposta totale Virgo +}\end{equation}

that, in the low frequencies limit ($\omega\rightarrow0$) is in perfect
agreement with the detector pattern of eq. (46) in \cite{key-13},
if one retains the first two terms of the expansion:

\begin{equation}
\begin{array}{c}
\tilde{H}^{+}(\omega\rightarrow0)=\frac{1}{2}(1+\cos^{2}\theta)\cos2\phi+\\
\\-\frac{1}{4}i\omega L\sin\theta[(\cos^{2}\theta+\sin2\phi\frac{1+\cos^{2}\theta}{2})](\cos\phi-\sin\phi).\end{array}\label{eq: risposta totale approssimata}\end{equation}

This result also confirms that the magnetic contribution to the variation
of the distance is an universal phenomenon because it has been obtained
starting from the full theory of gravitational waves in the TT gauge
(see also \cite{key-13}).

The same analysis can be now performed for the $\times$ polarization.
In this case, from eq. (\ref{eq: metrica TT totale}) the line element
is:

\begin{equation}
ds^{2}=-dt^{2}+dz^{2}+dx^{2}+dy^{2}+2h_{\times}(t+z)dxdy,\label{eq: metrica polarizzazione per}\end{equation}

and, by using eqs. (\ref{eq: rotazione}), (\ref{eq: rotazione 2})
and (\ref{eq: trasformazione metrica}), in the new rotated frame
the line element (\ref{eq: metrica polarizzazione per}) in the $u$
direction becomes:

\begin{equation}
ds^{2}=-dt^{2}+[1-2\cos\theta\cos\phi\sin\phi h_{\times}(t+u\sin\theta\cos\phi)]du^{2}.\label{eq: metrica per  lungo u}\end{equation}

Then the response function of the $u$ arm of the interferometer to
the $\times$ polarization is:

\begin{equation}
\Upsilon_{u}^{\times}(\omega)=\frac{-\cos\theta\cos\phi\sin\phi}{L}\tilde{H}_{u}(\omega,\theta,\phi),\label{eq: risposta per lungo u}\end{equation}

while the line element (\ref{eq: metrica polarizzazione per}) in
the $v$ direction becomes:

\begin{equation}
ds^{2}=-dt^{2}+[1+2\cos\theta\cos\phi\sin\phi h_{\times}(t+u\sin\theta\sin\phi)]dv^{2}\label{eq: metrica per  lungo v}\end{equation}

and the response function of the $v$ arm of the interferometer to
the $\times$ polarization is:

\begin{equation}
\Upsilon_{v}^{\times}(\omega)=\frac{\cos\theta\cos\phi\sin\phi}{L}\tilde{H}_{v}(\omega,\theta,\phi).\label{eq: risposta per lungo v}\end{equation}

Thus, the total frequency-dependent response function of an interferometer
to the $\times$ polarization is:

\begin{equation}
\tilde{H}^{\times}(\omega)=\frac{-\cos\theta\cos\phi\sin\phi}{L}[\tilde{H}_{u}(\omega,\theta,\phi)+\tilde{H}_{v}(\omega,\theta,\phi)]\label{eq: risposta totale Virgo per}\end{equation}

that, in the low frequencies limit ($\omega\rightarrow0$), is in
perfect agreement with the detector pattern of eq. (46) of \cite{key-13},
if one retains the first two terms of the expansion:

\begin{equation}
\tilde{H}^{\times}(\omega\rightarrow0)=-\cos\theta\sin2\phi-i\omega L\sin2\phi(\cos\phi+\sin\phi)\cos\theta.\label{eq: risposta totale approssimata 2}\end{equation}

The total lengths of the two arms in presence of the $\times$ polarization
and in the frequency domain are: \begin{equation}
\tilde{T}_{u}(\omega)=(\cos\theta\cos\phi\sin\phi)\tilde{H}_{u}(\omega,\theta,\phi)\tilde{h}_{\times}(\omega)+T.\label{eq: lunghezza u}\end{equation}
\begin{equation}
\tilde{T}_{v}(\omega)=(-\cos\theta\cos\phi\sin\phi)\tilde{H}_{v}(\omega,\theta,\phi)\tilde{h}_{\times}(\omega)+T,\label{eq: lunghezza v}\end{equation}

that also are particular cases of the more general equation (39) of
\cite{key-13}. The total low frequencies response functions of eqs.
(\ref{eq: risposta totale approssimata}) and (\ref{eq: risposta totale approssimata 2})
are more accurate than the ones of \cite{key-24,key-25}, because
our equations include the {}``magnetic'' contribution (see also
\cite{key-13}).

Then, we have shown that a generalization of the analysis in \cite{key-12,key-16}
works in the computation of the response functions of interferometers
and that our results in the frequency domain are totally consistent
with the results of \cite{key-13}. Thus the obtained results confirm
the presence and importance of the so-called {}``magnetic'' components
of GWs and the fact that they have to be taken into account in the
context of the total response functions of interferometers for GWs
propagating from arbitrary directions. 

The absolute values of the total response functions of the Virgo interferometer
for the $+$ and $\times$ polarizations of GWs propagating from the
direction $\theta=\frac{\pi}{4}$ and $\phi=\frac{\pi}{3}$ are shown,
respectively, in figs. 11 and 12. The same response functions are
shown in figs. 13 and 14 for the LIGO interferometer. We can see from
the figures that at high frequencies the absolute values of the response
functions decrease with respect to the constant values of the low
frequencies approximation. Finally, the angular dependences of the
total response functions of the Virgo interferometer to the $+$ and
$\times$ polarizations for $f=100Hz$ are shown in figs. 15 and 16
. The same angular dependences are shown for the LIGO interferometer
in figs. 17 and 18. 

\begin{figure}
\includegraphics{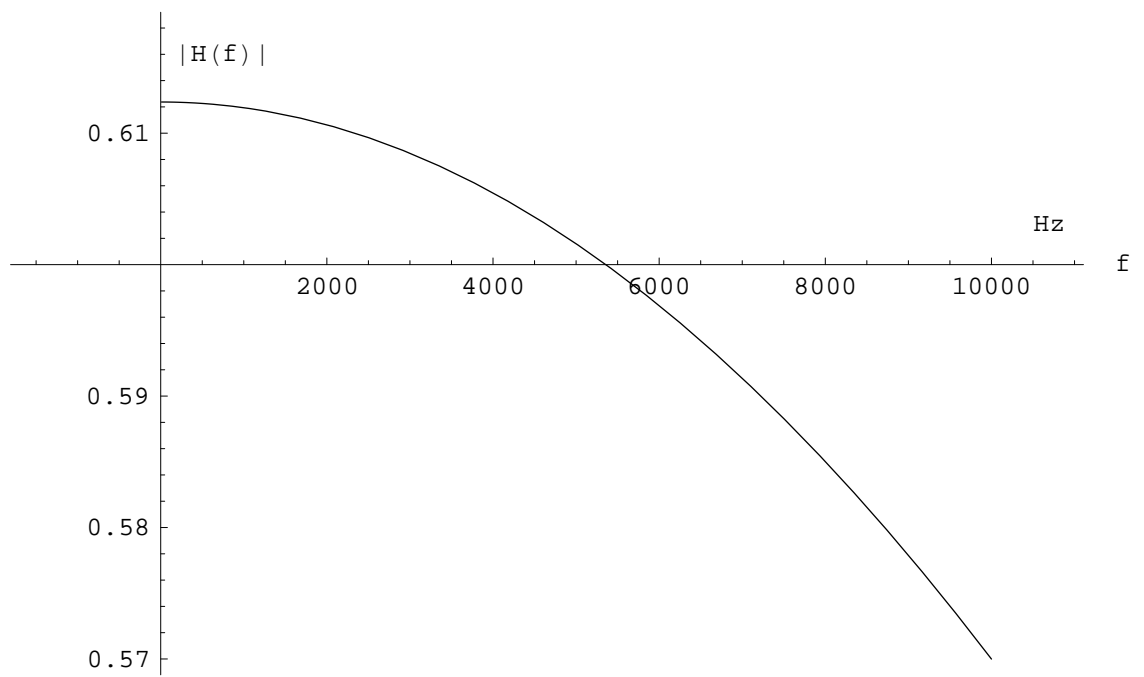}

\caption{the absolute value of the total response function of the Virgo interferometer
to the $+$ polarization for $\theta=\frac{\pi}{4}$ and $\phi=\frac{\pi}{3}$. }
\end{figure}
\begin{figure}
\includegraphics{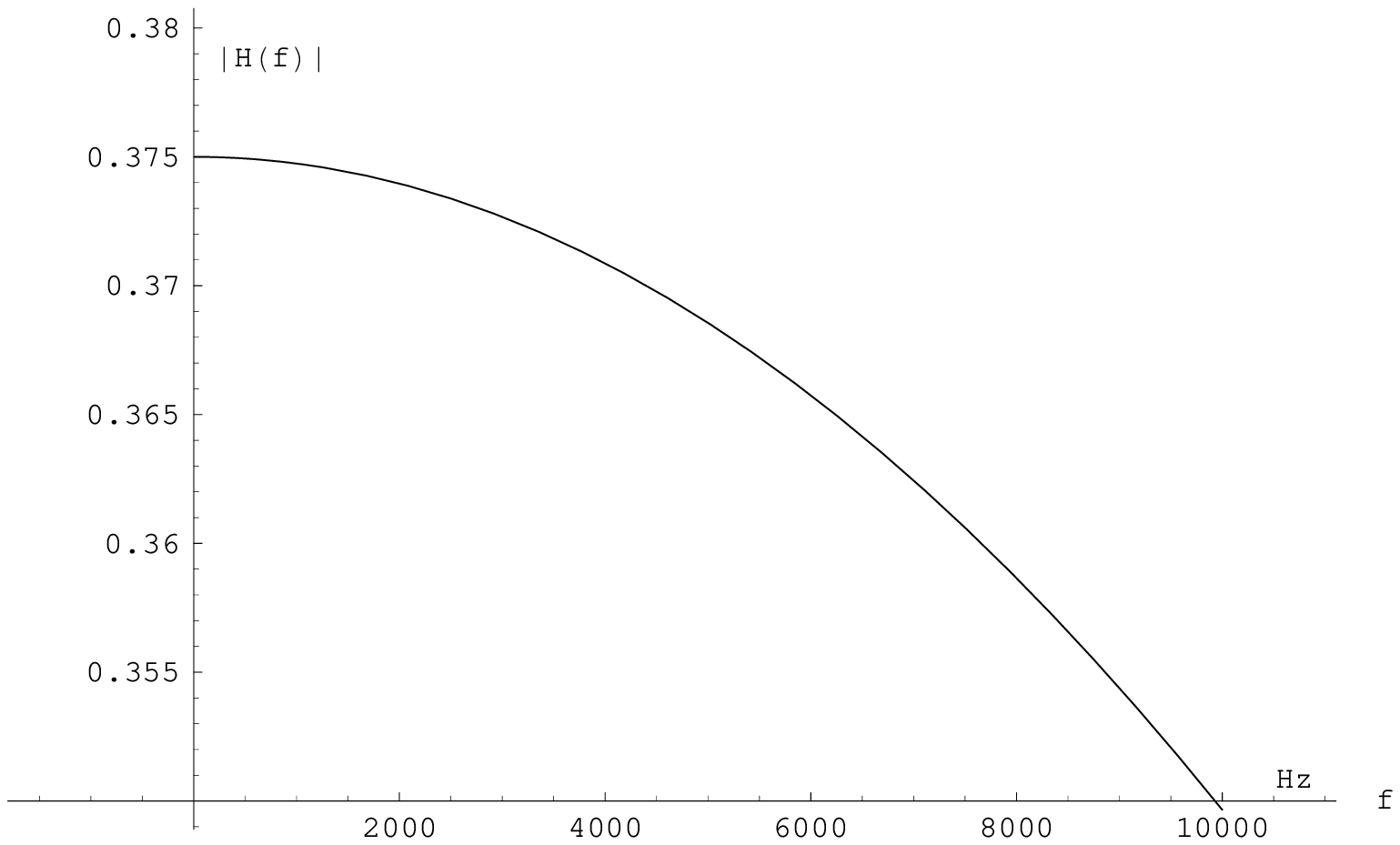}

\caption{the absolute value of the total response function of the Virgo interferometer
to the $\times$ polarization for $\theta=\frac{\pi}{4}$ and $\phi=\frac{\pi}{3}$. }
\end{figure}
\begin{figure}
\includegraphics{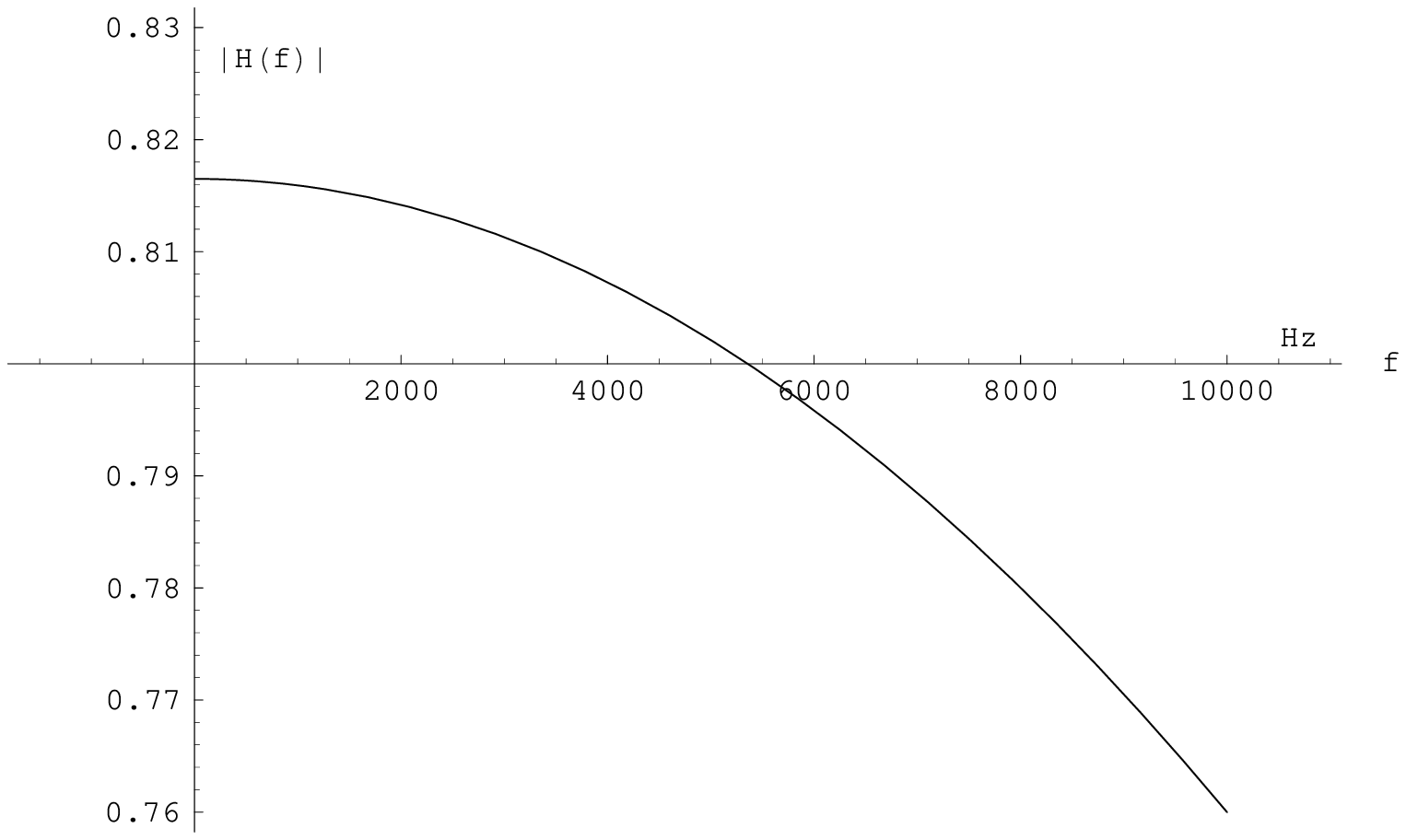}

\caption{the absolute value of the total response function of the LIGO interferometer
to the $+$ polarization for $\theta=\frac{\pi}{4}$ and $\phi=\frac{\pi}{3}$. }
\end{figure}
\begin{figure}
\includegraphics{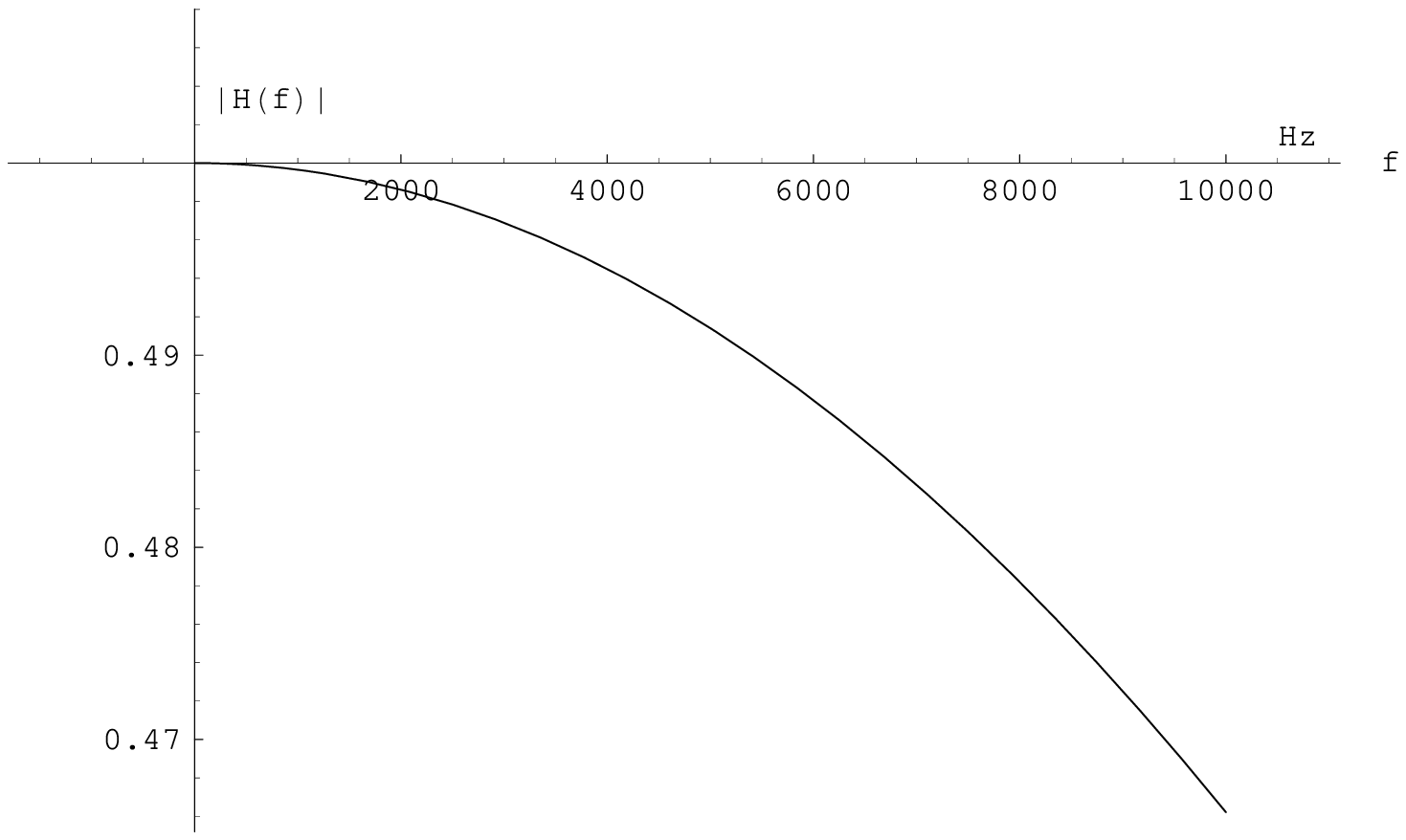}

\caption{the absolute value of the total response function of the LIGO interferometer
to the $\times$ polarization for $\theta=\frac{\pi}{4}$ and $\phi=\frac{\pi}{3}$. }
\end{figure}
\begin{figure}
\includegraphics{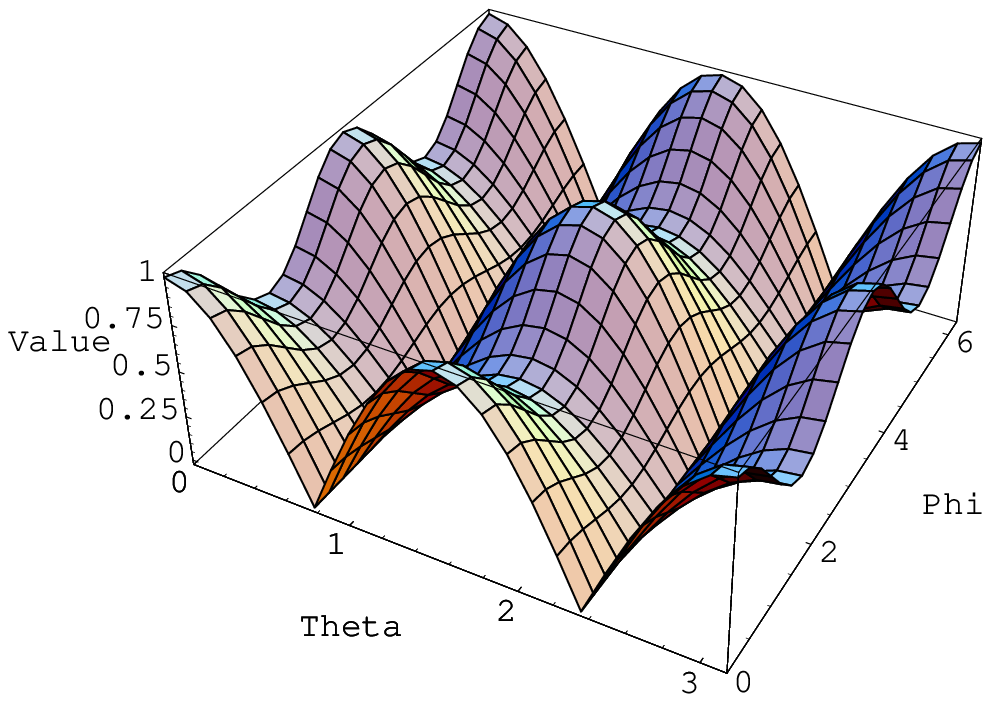}

\caption{the angular dependence of the total response function of the Virgo
interferometer to the $+$ polarization for $f=100Hz$}
\end{figure}
\begin{figure}
\includegraphics{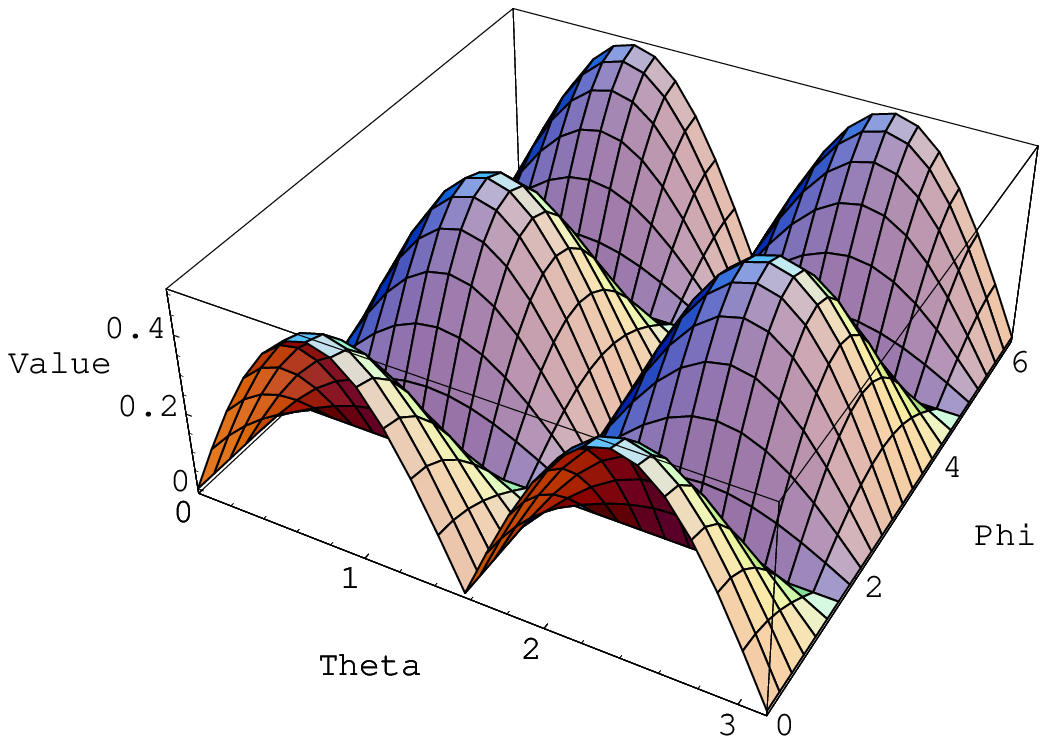}

\caption{the angular dependence of the total response function of the Virgo
interferometer to the $\times$ polarization for $f=100Hz$}
\end{figure}
\begin{figure}
\includegraphics{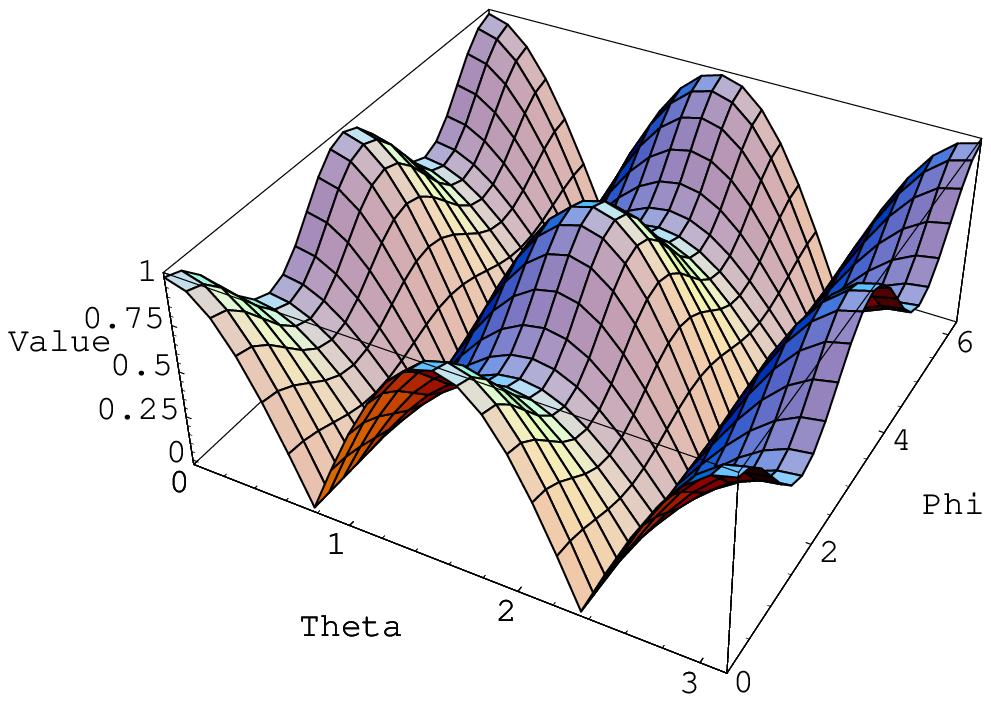}

\caption{the angular dependence of the total response function of the LIGO
interferometer to the $+$ polarization for $f=100Hz$}
\end{figure}
\begin{figure}
\includegraphics{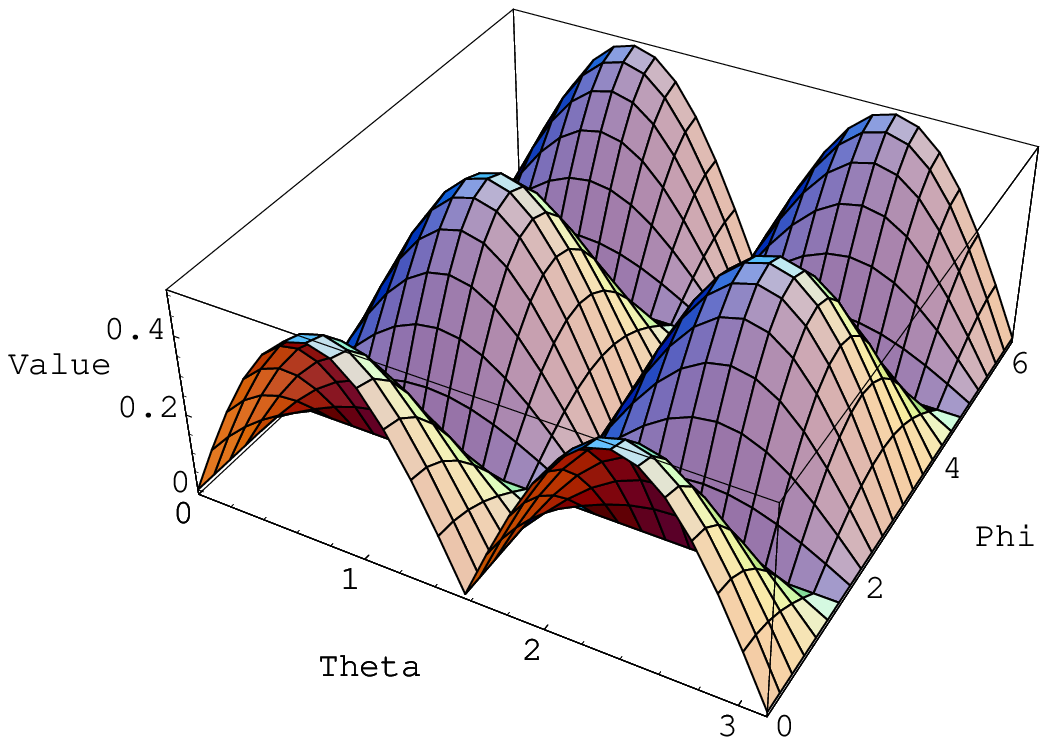}

\caption{the angular dependence of the total response function of the LIGO
interferometer to the $\times$ polarization for $f=100Hz$}
\end{figure}

\section{Conclusions}

In this paper detailed angular and frequency dependences of the response
functions for the magnetic components of GWs have been given in the
approximation of wavelength much larger than the linear dimensions
of the interferometer, with a specific application to the parameters
of the LIGO and Virgo interferometers. The presented results agree
with the work of \cite{key-13} in which it has been shown that the
identification of {}``electric'' and {}``magnetic'' contributions
is unambiguous in the long-wavelength approximation. At the end of
this paper the angular and frequency dependences of the total response
functions of the LIGO and Virgo interferometers have been given. In
the high-frequency regime the division on {}``electric'' and {}``magnetic''
components becomes ambiguous, thus the full theory of gravitational
waves has been used. The results of this work are consistent with
the ones of \cite{key-13} in this case too.

\section*{Acknowledgements}

I would like to thank Francesco Rubanu, Maria Felicia De Laurentis
and Giancarlo Cella for helpful suggestions and discussions during
my work. I thank the referee for its interest in my work and for precious
suggestions and comments that allowed to improve this paper and gave
to me a better knowledge of the physics of the {}``magnetic'' components
of GWs. The European Gravitational Observatory (EGO) consortium has
also to be thanked for the using of computing facilities.

\end{document}